\documentclass{aa}  

\usepackage{graphicx}
\usepackage{txfonts}
\usepackage{txfonts}
\usepackage{color}

\begin{document}

   \title{A new view on exoplanet transits:}
\subtitle{Transit of Venus described using three-dimensional solar atmosphere \textsc{Stagger}-grid simulations}
\titlerunning{A new view on exoplanet transits: Transit of Venus described using three-dimensional solar atmosphere}
  \author{A. Chiavassa \inst{1}, C. Pere\inst{1}, M. Faurobert\inst{1} G. Ricort\inst{1}, P. Tanga\inst{1}, Z. Magic\inst{2,3}, R. Collet\inst{4}, M. Asplund\inst{4}}
\authorrunning{A. Chiavassa et al.}
\institute{ Laboratoire Lagrange, UMR 7293, CNRS, Observatoire de la C\^ote d'Azur, Universit\'e de Nice Sophia-Antipolis, Nice, France \\
\email{andrea.chiavassa@oca.eu} \and 
Niels Bohr Institute, University of Copenhagen, Juliane Maries Vej 30, DK--2100 Copenhagen, Denmark  \and 
Centre for Star and Planet Formation, Natural History Museum of Denmark, University of Copenhagen, {\O}ster Voldgade 5-7, DK--1350 Copenhagen, Denmark
\and
 Research School of Astronomy $\&$ Astrophysics, Australian National University, Cotter Road, Weston ACT 2611, Australia}

 \date{...; ...}

 
  \abstract
    {An important benchmark for current observational techniques and theoretical modeling of exoplanet's atmosphere is the transit of Venus (ToV). Stellar activity and, in particular, convection-related surface structures, potentially cause fluctuations that can affect the transit light curves. Surface convection simulations can help the interpretation of ToV and also other transits outside our Solar System.}
    {We used realistic three-dimensional (3D) radiative hydrodynamical (RHD) simulation of the Sun from the \textsc{Stagger}-grid and synthetic images computed with the radiative transfer code {{\sc Optim3D}} to provide predictions for the transit of Venus (ToV) in 2004 observed by the satellite \textit{ACRIMSAT}.}
{We computed intensity maps from RHD simulation of the Sun and produced synthetic stellar disk image as a observer would see, accounting for the centre-to-limb variations. The contribution of the solar granulation has been considered during the ToV. We computed the light curve and compared it to the \textit{ACRIMSAT} observations and also to the light curves obtained with solar surface representations carried out using radial profiles with different limb-darkening laws. We also applied the same spherical tile imaging method used for RHD simulation to the observations of center-to-limb Sun granulation with \textit{HINODE}.}
    {We managed to explain \textit{ACRIMSAT} observations of 2004 ToV and showed that the granulation pattern causes fluctuations in the transit light curve. We compared different limb-darkening models to the RHD simulation and evaluated the contribution of the granulation to the ToV. We showed that the granulation pattern can partially explain the observed discrepancies between models and data. Moreover, we found that the overall agreement between real and RHD solar granulation is good, either in term of depth or Ingress/Egress slopes of the transit curve. This confirms that the limb-darkening and the granulation pattern simulated in 3D RHD Sun represent well what is imaged by \textit{HINODE}. In the end, we found that the Venus's aureole contribution during ToV is $\sim10^{-6}$ times less intense than the solar photosphere, and thus, accurate measurements of this phenomena are extremely challenging.}
  {The prospects for planet detection and characterization with transiting methods are excellent with access to large a amount of data for stars. Being able to explain consistently the data of 2004 ToV is a new step forward for 3D RHD simulations that are becoming essential for the detection and characterization of exoplanets. They show that the granulation have to be considered as an intrinsic incertitude, due to the stellar variability, on precise measurements of exoplanet transits of, most likely, planets with small diameters. In this context, the importance of a comprehensive knowledge of the host star, including the detailed study of the stellar surface convection is determinant.}
  

   \maketitle
%

\section{Introduction}

The detection of exoplanet orbiting around stars is now a routine with more that 1500 confirmed planets and about 3400 candidates \citep[as in fall 2014 from http://exoplanets.org, ][]{2011PASP..123..412W}. The study of exoplanet atmospheres can generally be classified into two main categories: (i) the measurement of time-varying signals, which include transits, secondary eclipses, and phase curves, and (ii) the spatially resolved imaging and spectroscopy.  These techniques allow to explore the property of exoplanet's atmospheres and understand the diversity of chemical compositions of exoplanets, their atmospheric processes, internal structures, and formation conditions \citep{2014arXiv1402.1169M}. 

A transit event occurs when an (or more) exoplanet move across the face of host star, hiding a small part of it, as seen by an observer at some particular vantage point. The exoplanet blocks some of the starlight during the transit and creates a periodic dip in the brightness of the star. The amount of light reduction is typically $\sim1\%$, 0.1$\%$, and, 0.01$\%$ for Jupiter-, Neptune- and Earth-like planets, respectively \citep{1984Icar...58..121B}. This depends on the star-planet size ratio and on the duration of the transit, which, in turn, depends on the planet's distance from the host star and on the stellar mass. This method is mostly sensitive to large, short-period exoplanets, which display the largest atmospheric signatures and make them popular targets for time-varying characterization studies \citep{2014arXiv1402.1169M}. During both the primary and secondary transits, the planet-star radius ratio can be deduced, and, for sufficiently bright stars, also the exoplanet's mass can be determined from the host star's radial velocity semi-amplitude \citep{2012A&A...538A...4M}. If the mass and/or radius of the exoplanet is good enough, also the atmospheric density can be predicted giving hints to the  current understanding of exoplanet formation processes. Even more information can be retrieved by measuring the depth of the transit: the radiation coming from the host star passes through the exoplanet's atmosphere and it is absorbed to different degrees at different wavelengths. This wavelength-dependent depth of the transmission spectrum is directly linked to the absorption features imprinted on starlight transmitted through the exoplanet's atmosphere \citep{2000ApJ...537..916S,2001ApJ...553.1006B}.\\
However, the transit method has several drawbacks: (i) to be observed, the exoplanet must cross directly the line-of-sight from the Earth (i.e., the star-planet system must be edge-on with respect to the observer) and this is the case only for a small minority of exoplanets. (ii) Another problem concerns the inter-transit duration that can last months to years, reducing drastically the number of detectable exoplanets. (iii) There is also a chance to observe the alignment with a background eclipsing binary that cause blends difficult to disentangle \citep{2011ApJ...727...24T}. (iv) A last potential complication is the host star surface inhomogeneities caused by the presence of stellar granulation \citep[e.g.,][]{2014A&A...567A.115C} or the magnetic starspots \citep[e.g., ][]{2009A&ARv..17..251S}.

A very important benchmark for current observational techniques and theoretical modeling of exoplanet's atmosphere is the transit of Venus (ToV). Venus is in many ways an Earth analogue and it's the closest planet transit we can see in our Solar System. ToV events are rare, as they occur in pairs 8 years apart, each pair separated by 121.5 or 105.5 years, alternating between descending node (June pairs: 1761/1769, 2004/2012,...), and ascending node (December pairs: 1631/1639, 1874/1882, 2117/ 2125,...). The 2004 and 2012 transits of Venus represent thus a crucial leap on human-life timescale, as the modern imaging technologies available allow for the first time a quantitative analysis of the atmospheric phenomena of the planet associated to its transit \citep{2012DPS....4450807T}. Spatial- and ground-based data have been collected and studied by different authors. \cite{2012DPS....4450807T} performed the first photometric analysis of the "aureole"\footnote{At the Ingress and Egress of ToV, the portion of the planet's disk outside the Solar photosphere has been repeatedly perceived as outlined by a thin and bright arc, called "aureole"} using various ground-based systems and extracted fundamental parameters of Venus's atmosphere. \cite{2012A&A...547A..22G} compared images of Venus during Ingress/Egress to 2004 data and provided guidelines for the investigation of the planet's upper haze from vertically-unresolved photometric measurements. \cite{2012A&A...537L...2E} provided a theoretical transmission spectrum of the atmosphere of Venus dominated by carbon dioxide absorption and droplets of sulfuric acid composing an upper haze layer (features not expected for a Earth-like exoplanet) that could be tested with spectroscopic observations of 2012 transit. 

In this work, we present the transit light curve predictions obtained from 3D surface convection simulation of the Sun and compare them to the ToV occurred in 2004. Furthermore, we compare 3D predictions to limb-darkening models of the Sun, largely used in the planetary community, as well as to the real observations of the granulation pattern on the Sun. The aim of this work is to test,  whether the 3D  surface convection approach to the transit technique gives reliable results for the benchmark case of the ToV, which also is important for the interpretation of other transits outside our Solar System.


\section{From the hydrodynamical simulation of the Sun to spherical tile imaging}

The Sun is a natural reference for studying other stars and solar models are essential for many studies in stellar and planetary astrophysics. Many of the observable phenomena occurring on the surface of the Sun are intimately linked to convection. Moreover, the atmospheric temperature stratification in the optical thin region, where the emerging flux form, is also affected by the interaction between radiative and convective energy transport. To account for all these aspects, it is important to use realistic 3D radiative hydrodynamical (RHD) simulations of convection \citep{1998ApJ...499..914S, 2012JCoPh.231..919F, 2004ApJ...610L.137C,2004A&A...421..741V}. RHD simulations of the Sun are widely used and compared in detail with observations by the astrophysical community using different numerical codes \citep[e.g., ][]{2009LRSP....6....2N,asplund09,2011SoPh..268..255C,2013A&A...554A.118P}.

\subsection{Stellar model atmospheres}

We used the solar simulation from the \textsc{Stagger}-grid of realistic 3D RHD simulations of stellar convection for cool stars \citep{2013A&A...557A..26M}. This grid is computed using \textsc{Stagger}-code (originally developed by Nordlund $\&$ Galsgaard 1995\footnote{http://www.astro.ku.dk/$\sim$kg/Papers/MHD\_code.ps.gz}, and continuously improved over the years by its user community). In a cartesian box located at the optical surface (i.e., $\tau\sim1$), the code solves the time-dependent equations for conservation of mass, momentum, and energy coupled to an realistic treatment of the radiative transfer.  The simulation domains are chosen large enough to cover at least ten pressure scale heights vertically and to allow for about ten granules to develop at the surface, moreover there are periodic boundary conditions horizontally and open boundaries vertically. At the bottom of the simulation, the inflows have a constant entropy and pressure.  The simulations employ realistic input physics: (i) the equation of state is an updated version of the one by \cite{1988ApJ...331..815M}; (ii) the radiative transfer is calculated for a large number over wavelength points merged into 12 opacity bins \citep{1982A&A...107....1N,2000ApJ...536..465S} with opacities including continuous absorption and scattering coefficients from \cite{2010A&A...517A..49H} and line opacities as described from \cite{2008A&A...486..951G}, in turn based on the VALD-2 database \citep{2001ASPC..223..878S} of atomic lines. The abundances employed in the computation of the Sun simulation are the latest chemical composition by \cite{asplund09}.

\begin{table*}
\centering
\begin{minipage}[t]{\textwidth}
\caption{3D simulation of the Sun from \textsc{Stagger}-grid used in this work.}             
\label{simus}      
\centering                          
\renewcommand{\footnoterule}{} 
\begin{tabular}{c c c c c c c c}        
\hline\hline                 
$<T_{\rm{eff}}>$\footnote{Horizontally and temporal average of the emergent effective temperatures from \cite{2013A&A...557A..26M}} & [Fe/H]  & $\log g$ & $x,y,z$-dimensions & $x,y,z$-resolution   & $\rm{M}_{\star}$ & $\rm{R}_{\star}$ & Number of tiles \footnote{$N_{\rm{tile}} = \frac{\pi \cdot \rm{R}_\odot}{x,y\rm{-dimension}}$} \\
$[\rm{K}]$ & & [cgs]  & [Mm]  & [grid points]   & [$\rm{M}_\odot$] & [$\rm{R}_\odot$] & over the diameter\\
\hline
5768.51 (Sun) & 0.0 & 4.4 &  7.7$\times$7.7$\times$5.2 & 240$\times$240$\times$240 & 1.0 & 1.0 & 286\\
\hline\hline                          
\end{tabular}
\end{minipage}
\end{table*}

\subsection{Three-dimensional radiative transfer}

The computation of the monochromatic emerging intensity is performed using the 3D pure local thermal equilibrium radiative transfer code \textsc{Optim3D} \citep{2009A&A...506.1351C} and the snapshots of the RHD simulation of the Sun (see Table~\ref{simus}). The radiative transfer equation is solved monochromatically using pretabulated extinction coefficients as a function of temperature, density, and wavelength based on the same extensive atomic and molecular continuum and line opacity data as the latest generation of MARCS models \citep{2008A&A...486..951G}. We assume zero microturbulence and model the non-thermal Doppler broadening of spectral lines using only the self-consistent velocity fields present in the 3D simulations. We employed the same chemical composition of the 3D RHD simulations \citep{asplund09}. The wavelength interval computed is $6684.0\pm 0.1$ $\AA$. The detailed methods used in the code are explained in \cite{2009A&A...506.1351C}.

\subsection{Spherical tile imaging}\label{tilingsection}

The computational domain of each simulation represents only a small portion of the stellar surface \citep[see, e.g., the central panel of Fig.~\ref{hinodeimages2} and Fig~1 in][]{2012A&A...540A...5C}. To obtain an image of the whole solar disk  accounting for limb-darkened effects, we employ the same tiling method explained in \cite{2010A&A...524A..93C} and used also in \cite{2012A&A...540A...5C,2014A&A...567A.115C}. \textsc{Optim3D} has been used to compute intensity maps from the solar simulation for different inclinations with respect to the vertical, $\mu{\equiv}\cos(\theta)$=[1.0000, 0.9890, 0.9780, 0.9460, 0.9130, 0.8610, 0.8090, 0.7390, 0.6690, 0.5840, 0.5000, 0.4040, 0.3090, 0.2060, 0.1040], with $\theta$ being the angle with respect to the line-of-sight (vertical axis), and for 30 representative snapshots of the simulation adequately spaced apart so as to capture several convective turnovers.

\begin{figure}
   \centering
   \begin{tabular}{c}  
   			\includegraphics[width=0.99\hsize]{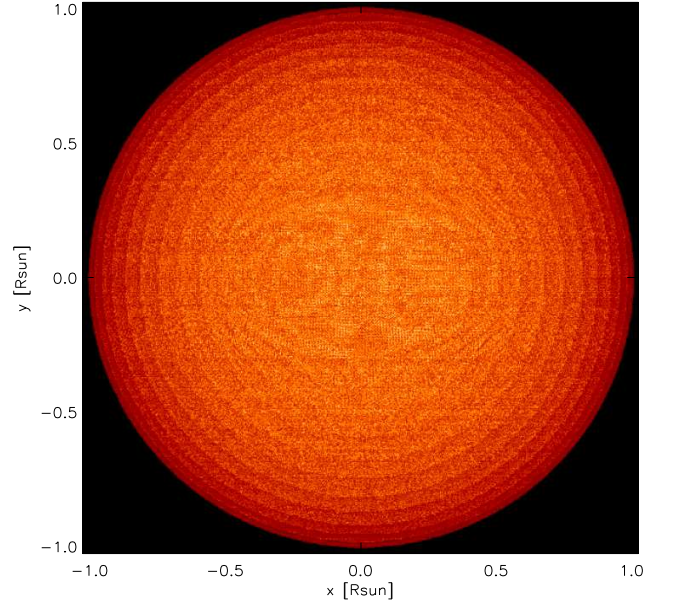}   
        \end{tabular}
      \caption{Synthetic solar disk image computed at $6684.0\pm 0.1$ $\AA$ of the RHD simulation of Table~\ref{simus}. The intensity range is [$0.0$--$3.79\times10^6$\,erg\,cm$^{-2}$\,s$^{-1}$\,{\AA}$^{-1}$]. We generated 50 different synthetic solar disk images to account for granulation changes with respect to time.}
        \label{starnoplanet}
   \end{figure}

\begin{figure}
   \centering
   \begin{tabular}{c}
               \includegraphics[width=0.98\hsize]{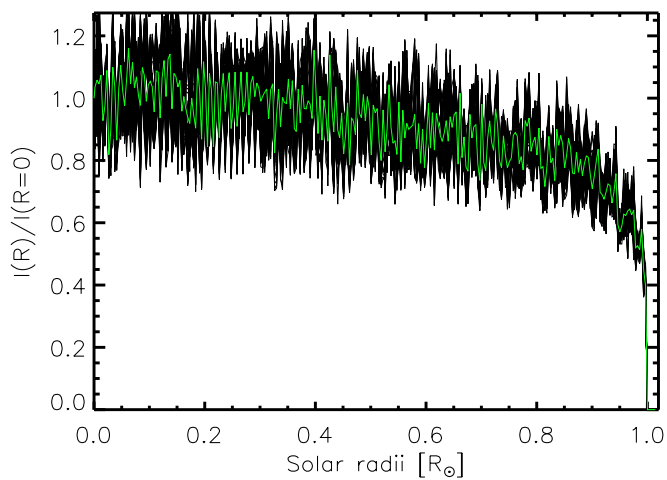}
        \end{tabular}
      \caption{Cut at x=0, y>0 for 50 different synthetic solar disk images (one example in reported Fig.~\ref{starnoplanet}). The intensity profiles are normalized to the intensity at the disk center (R=0.0). The green line is the RHD <3D> average profile.}
        \label{multigranulation}
   \end{figure}

We then used the synthetic images, chosen randomly among the snapshots in the time-series, to map onto spherical surfaces accounting for distortions especially at high latitudes and longitudes cropping the square-shaped intensity maps when defining the spherical tiles. The computed value of the $\theta$-angle used to generate each map depended on the position (longitude and latitude) of the tile on the sphere and was linearly interpolated among the inclination angles. The random choice of snapshots avoid the assumption of periodic boundary conditions and the resulting tiled spherical surface displays an artifactual periodic granulation pattern (Fig.~\ref{starnoplanet}). The total number of tiles ($N_{\rm{tile}}$) needed to cover half a circumference from side to side on the sphere is $N_{\rm{tile}} = \frac{\pi \cdot \rm{R}_\odot}{x,y\rm{-dimension}} = 286$, where $\rm{R}_\odot$ (transformed in Mm) and the  $x,y\rm{-dimension}$ come from Table~\ref{simus}. We generated 50 different synthetic solar disk images. Since the number of tiles necessary to cover the sphere is larger than the total number of representative snapshots of RHD simulation, some tiles may randomly appears several times on the solar disk at different inclinations. Therefore, the 50 synthetic images are not completely independent, but we assume that they are a good enough statistical representation to estimate the granulation changes with respect to time during the ToV. Fig.~\ref{multigranulation} displays the fluctuation of the granulation structures for a particular cut in the synthetic disk images. The solar disk intensity fluctuates at the scale of the tile and the maximum intensity varies between [3.74--3.87] $\times10^6$\,erg\,cm$^{-2}$\,s$^{-1}$\,{\AA}$^{-1}$.

\section{Modeling the transit of Venus of 2004}

\begin{figure}
   \centering
   \begin{tabular}{c}  
   			\includegraphics[width=1\hsize]{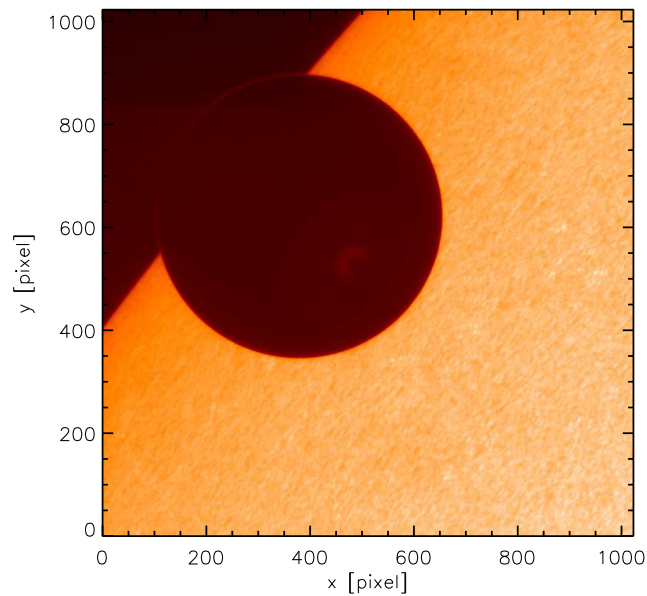} \\
			\includegraphics[width=0.7\hsize]{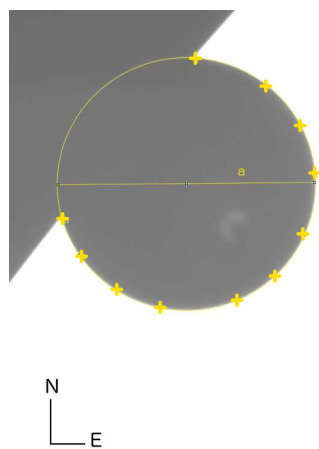} \\
			\includegraphics[width=0.9\hsize]{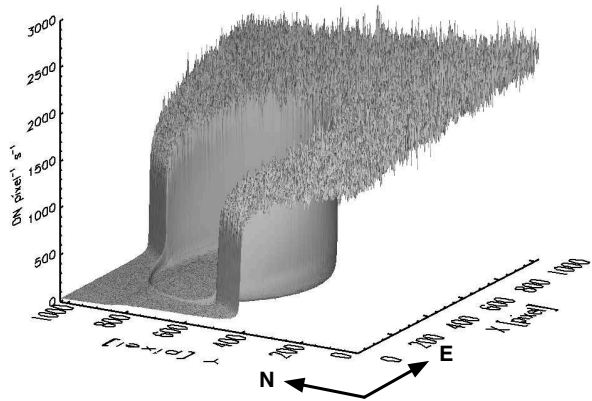}   
        \end{tabular}
         \caption{\emph{Top panel: } Image ($1024\times1024$ pixels) of Venus (dark circular subaperture) transiting the Sun (brighter areas) in June 5th 2012 as seen by \textit{HINODE} SOT/SP spectropolarimeter (total intensity Stoke parameter $I$) in the red filter centered at $6684.0\pm 0.1$ $\AA$ (Red filter). The intensity range is $0.$--$2.7\times10^3$\,DN\,pixel$^{-1}$\,s$^{-1}$ \citep[where DN is data numbers, ][]{2013SoPh..283..579L}. The intensity plotted is the square root of the intensity to better display the "aureole". \emph{Central panel: } Same figure as above with the contrast highly increased to fit the radius of the planet, yellow circle, with the procedure described in the text ($R_{\rm{fit}}=274.91$ pixels). The center of the Sun is located in South-East direction. \emph{Bottom panel: }Three dimensional view of the images above. The aureole contribution is visible in the region just outside the solar disk.} 
        \label{hinode}
   \end{figure}    

June 8th 2004 was a very important and rare date for monitoring the first of two close by (2004 and 2012) transits of Venus visible from Earth, before 2117. \cite{2006ApJ...641..565S}, hereafter SPW2006, took the chance and obtained space-based solar irradiance measurements with \textit{ACRIMSAT} spacecraft. 

\subsection{Observations with ACRIMSAT}\label{acrimsection}

The satellite mission \textit{ACRIMSAT}\footnote{http://acrim.jpl.nasa.gov}, launched in 1999 and founded by NASA through the Earth Science Programs Office at Goddard Space Flight Center,  measured 0.0961$\%$ reduction in the solar intensity caused by the shadow of the Venus during his transit in 2004 (SPW2006). These measurements were taken with the Active Cavity Radiometer Irradiance Monitor (\textit{ACRIM} 3) instrument \citep{2003GeoRL..30.1199W}. \textit{ACRIM} 3 was designed to provide accurate, highly precise, and traceable radiometry over decadal timescales, to detect changes in the total energy received from the Sun by the Earth. \textit{ACRIM} 3 provides 32 second exposure shutter measurements characterized by open (observations) and closed (calibration) measurements. The measurement precision is $0.01\%$ (SPW2006). 

\textit{ACRIMSAT} cannot observe the Sun continuously because the line-of-sight is occulted by the Earth during its orbit around the Sun. In SPW2006, these periodic interruptions were determined for time intervals spanning the ToV using a definitive a posteriori orbital ephemeris derived from a contemporaneous epochal satellite element set provided by the North American Aerospace Defense Command. SPW2006 reports (Figure 2 in SPW2006) the apparent path of the center of Venus as seen from \textit{ACRIMSAT} with respect to the heliocenter as the planet crossed the face of the Sun. \textit{ACRIMSAT}'s observations are interrupted for $\sim$30 minutes by Earth occultation during each of its $\sim$100 minute orbits. The duration of 2004 ToV is $\sim$5.5 hours.\\
In this work, we use the trajectory coordinates of Venus extracted from SPW2006 to simulate the passage of the planet on the synthetic solar disk of Fig.~\ref{starnoplanet}.

\subsection{Evaluation of Venus's aureole with \textit{HINODE}}

Telescope images, with instrument typically up to 15 to 20 cm, report cusp extension tending to transform the thin crescent of Venus into a ring of light since the mid 18th century \citep[for detailed historical reviews see][]{1969epa..book.....L,edson}. While these phenomena can be ascribed to light diffusion, during transits the contribution of refraction (3-4 orders of magnitude stronger) shows up. Traditionally, its discovery is attributed to Lomonosov \citep[who lived between 1711 and 1765; ][]{2005tvnv.conf..209M}. As the ToV are rare, also the data concerning the aureole are sparse. However, thanks to the coordinated giant leap in the observation of 2004 ToV, as well as the modern technologies allowed a quantitative analysis of the atmospheric phenomena associated to the ToV \citep{2012DPS....4450807T}. The authors performed for the first photometric analysis and obtained spatially resolved data providing measurements of the aureole flux as a function of Venus's latitude along the limb. 

The aureole is a distorted (strongly flattened) image of a portion of the solar photosphere. As such the surface brightness of the photosphere is the same in the aureole. For this reason, although the aureole is very thin (< 40 km above the cloud deck) it is clear the overall contribution to the transit photometry deserves a specific evaluation. To do this, we needed high spatial resolution images of Venus during a transit, and, only \textit{HINODE} spacecraft could provide that. Unfortunately, the satellite was launched after 2004 and thus we used observations carried out during ToV of 2012, assuming that there were no changes in Venus's atmosphere between 2004 and 2012.\\
\textit{HINODE} \citep{2007SoPh..243....3K} is a joint mission between the space agencies of Japan, United States, Europe, and United Kingdom launched in 2006, and since then, the spectropolarimeter SOT/SP \citep{2001ASPC..236...33L,2008SoPh..249..167T} allows the solar community to perform spectropolarimetry of the solar photosphere under extremely stable conditions. SOT/SP provides 0.32 arcsec angular resolution measurements of the vector magnetic fields on the Sun through precise spectropolarimetry of solar spectral lines with a spatial resolution of 0.32 arcsec and photometric accuracy of $10^{-3}$ \citep{2008SoPh..249..233I}. \\

We obtained \textit{HINODE} observations of June 5th 2012 ToV from through the \textit{HINODE} data on-line archive. We used SOT/SP Stokes $I$ images (i.e., the total intensity measured) in the red filter ($6684.0\pm 0.1$ $\AA$). The data are reduced and calibrated for all the slit positions (i.e., corresponding to "level1" data). The solar diameter in that period was 31.7\arcmin ($\sim$951.1\arcsec for the radius), the geometrical distance at the Sun surface, seen under one arcsecond was $R_{\odot \rm{SOHO}}/951.1\times0.109=732.1$ km \citep[with $R_{\odot \rm{SOHO}} =696342$ km measured by SOHO spacecraft,][]{2012ApJ...750..135E}.
The spatial resolution of the image is 1024x1024 pixels (Fig.~\ref{hinode}) and 1 pixel = 0.109\arcsec, the size of each image is then $111.61\times111.61\arcsec$ or $81.71\times81.71$ Mm, with 1 pixel equal to 79.80 km.\\

\begin{figure}
   \centering
   \begin{tabular}{cc}
         \includegraphics[width=0.91\hsize]{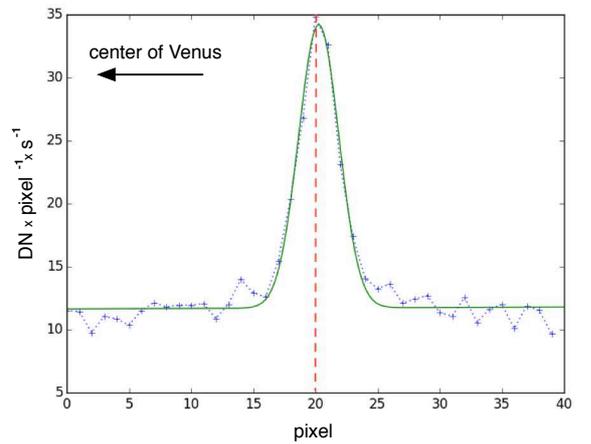}
        \end{tabular}
      \caption{Representative radial intensity profiles within the ring region ($R_{\rm{fit}}\pm20m$ pixels) in the image of Fig.~\ref{hinode}, which includes the aureole contribution. The blue dotted curve corresponds to the observations, while the green solid line to the best fit (see text for details). The vertical red dashed line is the location of Venus's radius.}
        \label{aureola}
   \end{figure}

The contribution from the aureole was computed as follows:

\begin{enumerate}
\item The dark circular subaperture corresponding to Venus in Fig.~\ref{hinode} (central panel) was fitted using Newton-based fit method \citep{Pratt:1987:DLF:37402.37420}. We selected the points where approximatively resides Venus's radius (yellow numbered crosses in the figure). The fitting method returned the center of the planet as well as its radius (or impact parameter), $a$ in the figure, with a Venus radius $R_{\rm{fit}}$=274.91 pixels.
\item The position of the center and the radius known, we extracted a ring of 40 pixels in the outer regions of Venus within the region $R_{\rm{fit}}\pm20$ pixels.
\item We extracted $215$ radial intensity profiles within the ring region (a representative example is reported in Fig.~\ref{aureola}).
\item We fitted the aureole contribution using a gaussian law (3 free parameters) for the peak and a straight line equation (2 free parameters) for the background. Then we integrated over the radius the intensity subtended by the gaussian fitting curve for each of the 215 radial profiles. We obtained the radial contribution of the aureole.
\item We repeated this procedure for 50 other images during the Ingress and Egress of ToV with similar results.
\end{enumerate}

Fig.~\ref{aureola} shows the typical signal of the aureole in the ring region selected close to Venus's radius. The peak is significant and much larger than the background signal (always lower than $\sim$20 DN\,pixel$^{-1}$\,s$^{-1}$). The contribution of the aureole corresponding to the 215 radial intensity profiles is in the range $I_{\rm{aureole}}=[0.4-6.0] \times 10^{5}$ DN\,pixel$^{-1}$\,s$^{-1}$ with an average value of $<I_{\rm{aureole}}>=1.2 \times 10^{5}$ DN\,pixel$^{-1}$\,s$^{-1}$ (Table~\ref{aureolatable}). \\
To relate the aureole intensity to the full Sun, we use the 3D synthetic solar disk image of Fig.~\ref{starnoplanet}. First, we define $I_{\rm{Hinode}}$ as the total emerging intensity from Fig~\ref{hinode} (top panel), $I_{\rm{synth global}}$ as the total emerging intensity from the synthetic Sun of Fig.~\ref{starnoplanet}, and $I_{\rm{synth cut}}$ as intensity emerging from a cut in the synthetic disk of Fig.~\ref{starnoplanet} corresponding to the same size and shape of HINODE observation of Fig~\ref{hinode}. Then the relative intensity due to the aureole is $I_{\rm{ratio}}=I_{\rm{aureole}} \times \left(I_{\rm{synth cut}}/I_{\rm{synth global}}\right)\times \left(1/I_{\rm{Hinode}}\right)$, knowing that $\left(I_{\rm{synth cut}}/I_{\rm{synth global}}\right)=0.0118$. Table~\ref{aureolatable} reports values of the order of $10^{-6}$, far too weak to be detected, also considering the precision level of \textit{ACRIMSAT} ($\sim10^{-5}$) and the height of the transit ($10^{-4}$, Fig.~\ref{transit}). We therefore do not consider the aureole in our modeling. 

\begin{table}[h!]
\centering
\caption{Integrated intensity contributions ($I_{\rm{aureole}}$) of maximum, minimum and average values from the aureole (Fig.~\ref{aureola}), together with the related ratio ($I_{\rm{ratio}}$) to the total emerging intensity from the synthetic Sun of Fig.~\ref{starnoplanet}.}            
\label{aureolatable}      
\centering                          
\renewcommand{\footnoterule}{} 
\begin{tabular}{c c c}        
\hline\hline                 
	    & $I_{\rm{aureole}}$ $\times 10^{5}$& $I_{\rm{ratio}}$$\times 10^{-6}$ \\
		    & [DN\,pixel$^{-1}$\,s$^{-1}$] &  \\
	    \hline
Average & 1.2 & 1.0 \\
Max & 6.0  & 5.4 \\
Min & 0.4  & 0.3 \\
\hline\hline                          
\end{tabular}
\end{table}

\subsection{Modeling the transit with 3D RHD simulation of the Sun}\label{planettransit}

\begin{figure}
   \centering
   \begin{tabular}{c}
    \vspace*{-40mm}
              \includegraphics[width=0.99\hsize]{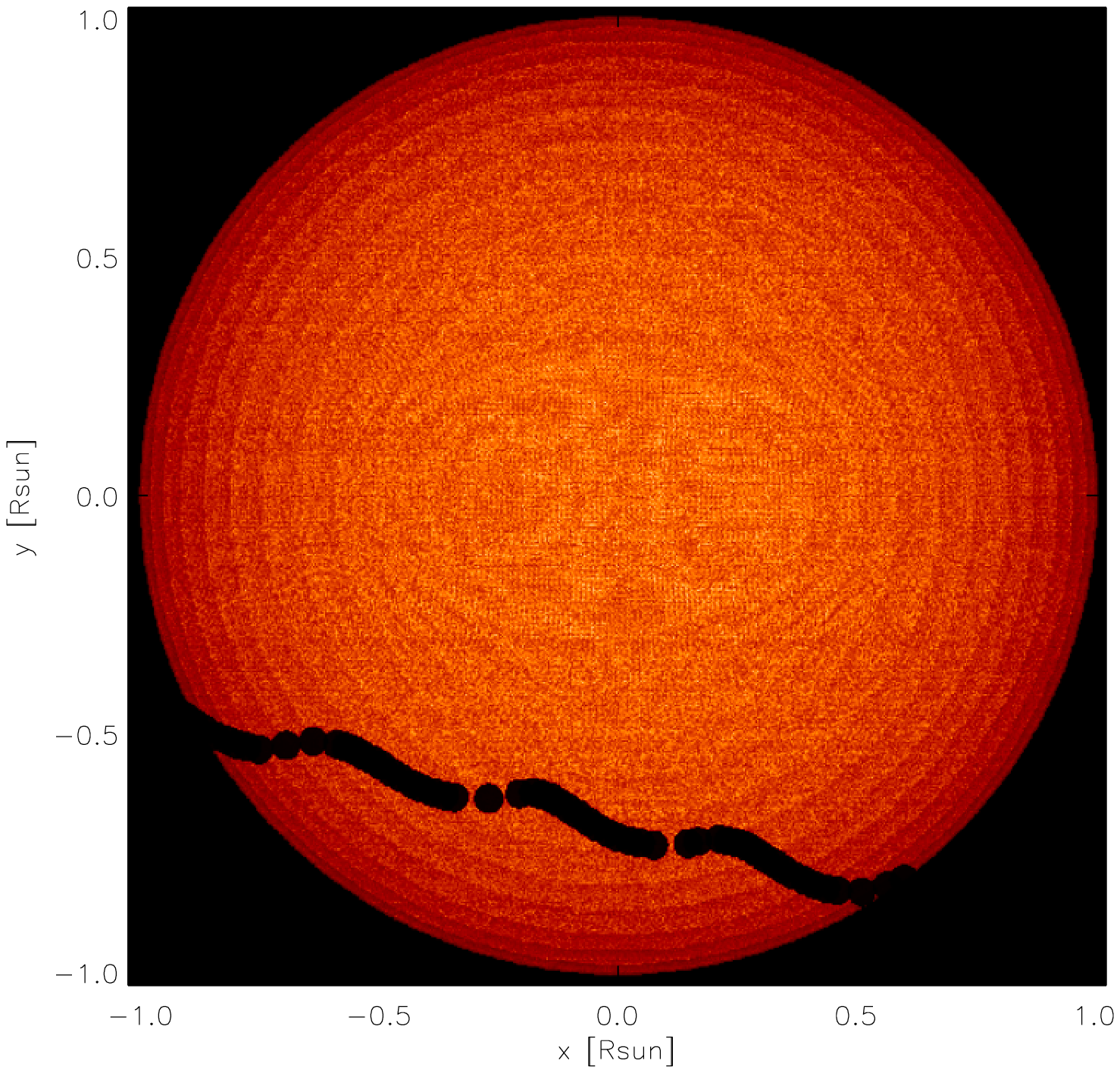}\\             
               \includegraphics[width=0.99\hsize]{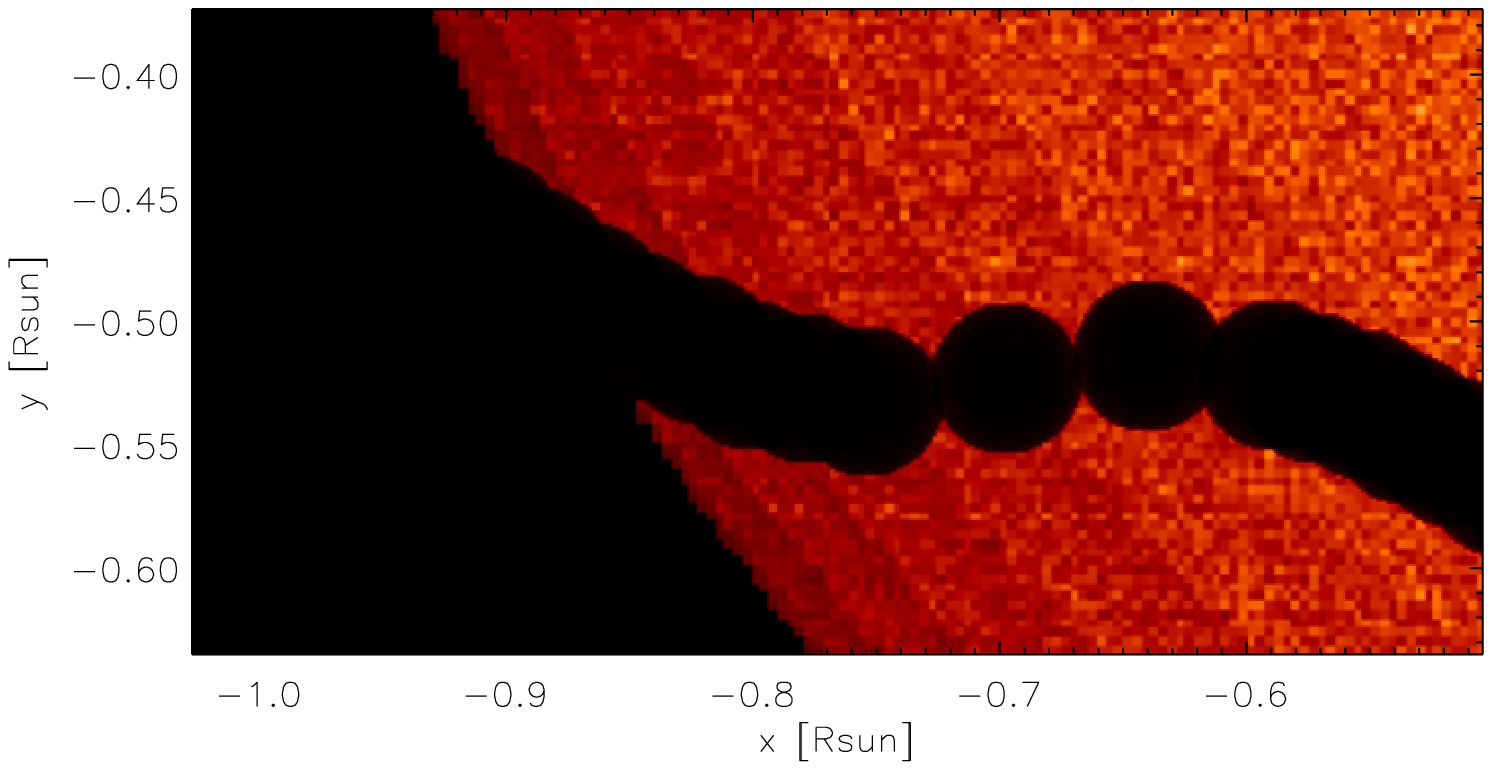}\\
        \end{tabular}
      \caption{\emph{Top panel: } Synthetic solar disk image in the visible of the RHD simulation with the different positions of ToV as seen from \textit{ACRIMSAT}. The computed wavelength is the same as in Fig.~\ref{starnoplanet}. The unusual apparent trajectory of Venus is induced by the spacecraft orbit (see text). \emph{Bottom panel: } enlargement from top panel.}
        \label{starplanet}
   \end{figure}

The observations give the transit parameters such as: the geometrically derived transit depth and contact times, the ratio between Venus's and Sun's radii, and the positions of Venus with respect to the heliocenter as a functions of time given the spacecraft and planetary orbits. In the following, we explain how we modeled the ToV using RHD simulation of the Sun as a background. The steps are:

\begin{enumerate}
\item we use the synthetic solar image (Fig.~\ref{starnoplanet}) as the background emitting source;
\item Venus is modeled by a disk of radius $R_{\rm{Venus}} = 0.0307 R_\odot \sim 8.5$ pixels (Fig.~\ref{starplanet}) with total intensity $0.0961\%$ of $I_{\rm{synth global}}$ (percentage flux decrement during 2004 transit, SPW2006);
\item the planet disk crosses the solar disk following the apparent trajectory of Venus as seen from \textit{ACRIMSAT} (see SPW2006 for more details and Fig.~\ref{starplanet}):
\item the intensity of the system is collected for every transiting step.
\end{enumerate}

The intensity map of Fig.~\ref{starplanet} is a statistical representation of the Sun at a particular moment during the transit. The apparent trajectory of Venus on the Sun is characterized by (see SPW2006 for more details): (i) a periodic spatial and temporal modulations in the location of Venus as it traverses the solar disk caused by the projection on the Sun of the shifting of the line-of-sight induced by the spacecraft orbit; (ii) a vertical (north-south) amplitude variations resulting from the near-polar orbit of the spacecraft; (iii) an horizontal (east-west) component manifesting in nonlinear spacings in the planetary position along its projected path in equal time intervals due to orbit plane not being on the line-of-sight direction to the Sun.

\begin{figure}
   \centering
   \begin{tabular}{c}
                       \includegraphics[width=0.99\hsize]{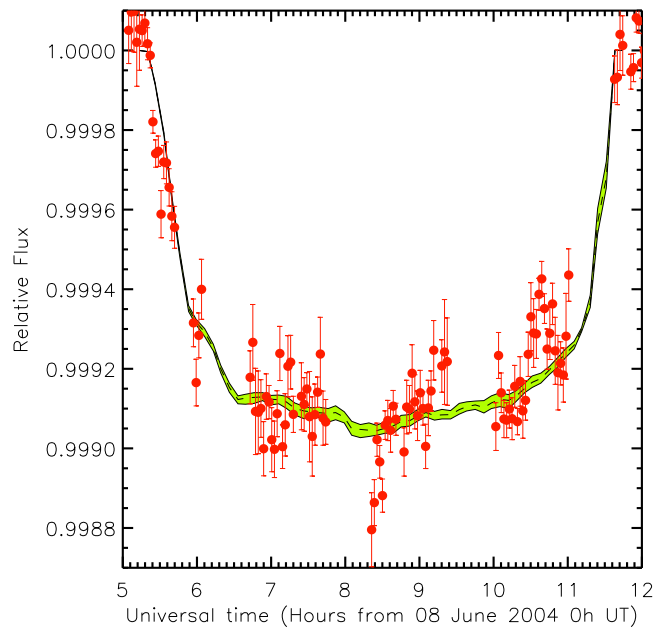}\\
                       \includegraphics[width=0.99\hsize]{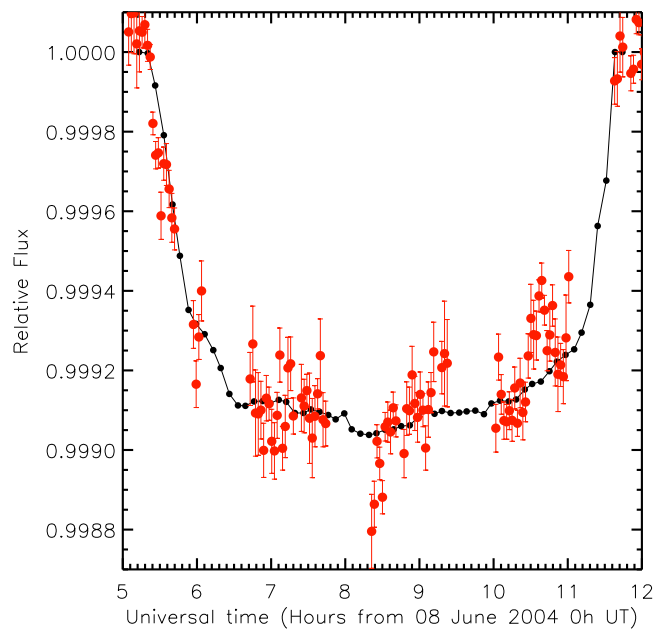}
        \end{tabular}
      \caption{\emph{Top panel: }ToV 2004 light curve (black dashed line) for the 3D RHD simulation of the Sun (Fig.~\ref{starplanet}) compared to the photometric observations \citep[red dots with error bars,][]{2006ApJ...641..565S} taken with \textit{ACRIM} 3 mounted on \textit{ACRIMSAT}. The observations have been normalized to 1, dividing the original data \cite[Figure 3 of ][]{2006ApJ...641..565S} by 1365.88 W/m$^2$. The light curve has been averaged (with the green shade denoting maximum and minimum values) over 50 different synthetic solar disk images to account for granulation changes with respect to time. \emph{Bottom panel: } same as above but for one particular realization among the 50 synthetic solar disk images.}
        \label{transit}
 \end{figure}

The temporal fluctuations is taken into account using the 50 different synthetic solar disk images computed. The resulting light curve from the ToV is reported in Fig.~\ref{transit}. The light curve has been averaged over 50 possible realizations of solar disk. The decrease of the Sun's light is due to the planet occultation and depends on the ratio between the respective size of Venus and the Sun as  well as on the brightness contrast. The magnitude of this intensity variation depends on the ratio of the size of Venus on the brightness contrast between the solar disk and the planet. As a result of the reflective spacecraft parallactic motion of Venus, the planet's apparent path crosses regions of the Sun with surface brightnesses very close to the stellar limb. This affects the slope of the light curve that is more shallow during Egress with respect to the Ingress. This aspect is very constraining for transit modeling and crucial to validate the RHD simulation used.
 
The data obtained in 2004 for the ToV (see Section~\ref{acrimsection}) has been provided to us by Glenn H. Schneider (University of Arizona, private communication). Unfortunately, 2012 ToV data are not helpful in our work because they had too large error bars. Venus occulted $0.0961\%$ of the total area of the Sun photosphere, and, \textit{ACRIM} 3 provided calibrated observations with gaps resulting from the Earth occultations during the satellite orbit. Fig.~\ref{transit} shows that the agreement between our RHD simulation and the data is very good. The root-mean-square deviation ($RMSD=\frac{\sum_{i=1}^{^n}\left(\hat{y}_i-yi\right)^2}{n}$; where $\hat{y}_i$ are the predicted values, ${y}_i$ the observed values, and $n$ the number of measurements) varies between [9.1-9.5]$\times10^{-5}$ for all the 50 possible realizations of solar granulation and is equal to 9.1$\times10^{-5}$ for the average profile plotted in Fig.~\ref{transit}. At the bottom of the light curve ($\sim$8.4h UT), there are some aperiodic variations. SPW2006 reported that this may result from intrinsic changes in the total solar irradiance over the same time interval or may arise as Venus occults isolated regions of the photosphere differing in local surface brightness (e.g., sunspots or smaller spatial scale localized features). Indeed, it is possible that magnetic starspots contaminate the transit signal \citep[e.g.,][]{2003ApJ...585L.147S,2009A&A...493..193L}. In the end, accounting for the granulation fluctuations allows to reduce the scatter on the observed points at about 7h, 9h, 11h UT (Fig.~\ref{transit}, bottom panel). \\
Our approach is statistical and aims not to fully explain the observed data but to show that our RHD simulation of the Sun is adapted (in terms of limb-darkening and emerging flux) to interpret such a data and that the granulation fluctuations have an important effect of the light curves during the transit. The evaluation of the granulation contribution is done in Section~\ref{limbdarkeningsection}.


\subsection{3D RHD granulation versus limb-darkening models}\label{limbdarkeningsection}

  \begin{figure}
   \centering
   \begin{tabular}{c}
               \includegraphics[width=0.98\hsize]{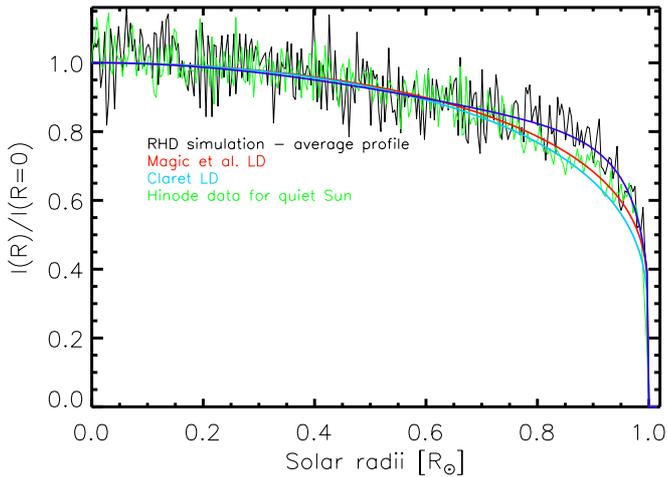}
        \end{tabular}
      \caption{Cut at x=0, y>0 in different synthetic solar disk images either built from \cite{2000A&A...363.1081C} (light blue curve), \cite{2014arXiv1403.3487M} (red curve), RHD <3D> profile (black curve, from Fig.~\ref{multigranulation}) and \textit{HINODE} data for the quiet sun (green curve, used in Section~\ref{hinodesection}). The intensity profiles are normalized to the intensity at the disk center (R=0).}
        \label{profilecomparison}
   \end{figure} 
   
In the literature, in general, the contribution of the star to the transit is approximated by a parametric representation of the radial limb-darkened surface profile. This approach neglects the surface inhomogenieties caused by the granulation pattern, the intensity brightness decrease is taken into account, depending on the limb-darkening law used. As a consequence, the slope of the light curve in the Ingress and Egress may be affected differently. In SPW2006, the authors used this method and examined the causes that could affect their deficiency in light curve fitting of the transit data. In their work, they argued that instrumental measurement errors and intrinsic solar variations compared to imperfections in the limb-darkening model itself may be the cause. In this context, RHD simulation of the Sun brings a new insight on the impact of the granulation thanks to the excellent matching showed in Fig.~\ref{transit}. The numerical box of a RHD simulation is rotated with a determined mu-angle (see Section~\ref{tilingsection}) before computing the emerging intensity. There is not a pre-adopted limb darkening function and the final result is the direct output of the emerging intensity at a particular mu-angle.

\begin{table}[h!]
\centering
\caption{Limb-darkening coefficients used in this work with the law $I_\mu/I_1=1-\sum_{k=1}^4a_k\left(1-\mu^{k/2}\right)$ \citep{2000A&A...363.1081C} and a star with solar parameters or for the <3D> profile of the RHD simulation (Fig.~\ref{multigranulation}, green line) or the intensity profile of \textit{HINODE data} (Fig.~\ref{profilecomparison}, green line).}            
\label{coeff}      
\centering                          
\renewcommand{\footnoterule}{} 
\begin{tabular}{c c c c c c c c}        
\hline\hline                 
Reference & $a_1$ & $a_2$ & $a_3$ & $a_4$  \\
Claret et al. 2000 & 0.5169 & -0.0211 & 0.6944 & -0.3892 \\
Magic et al. 2014 & 0.5850 & -0.1023 & 0.5371 &  -0.2580 \\
Fit RHD <3D> & -3.8699  &  13.3547  &  -15.4237   & 6.1910 \\
Fit \textit{HINODE} data & 18.32 & -38.49 & 36.05 & -12.20 \\
\hline\hline                          
\end{tabular}
\end{table}
 
We tested different limb-darkening laws represent the Sun. For this purpose, we built synthetic images representing the solar surface using radial profiles obtained with the limb-darkening law of  \citeauthor{2000A&A...363.1081C} \citeyear{2000A&A...363.1081C} (based on 1D model atmospheres of \citeauthor{1979ApJS...40....1K} \citeyear{1979ApJS...40....1K}): $I_\mu/I_1=1-\sum_{k=1}^4a_k\left(1-\mu^{k/2}\right)$, expressed as the variation in intensity with $\mu$-angle that is normalized to the disk-center ($I_\mu/I_1$). The limb-darkening coefficients for a star with the stellar parameters of the Sun and for the visible range are in Table~\ref{coeff}. The reader should note that \cite{2014arXiv1403.3487M} used the same RHD simulation of Table~\ref{simus} to extract their coefficients. Moreover, we used the law of \citeauthor{2000A&A...363.1081C} to fit the  average RHD profile $<3D>$ of Fig.~\ref{multigranulation} (green curve) and reported the coefficients in Table~\ref{coeff}. \\
The normalized intensity profiles of the different laws are reported in Fig.~\ref{profilecomparison}. 
 
 \begin{figure}
   \centering
   \begin{tabular}{c}
                       \includegraphics[width=0.99\hsize]{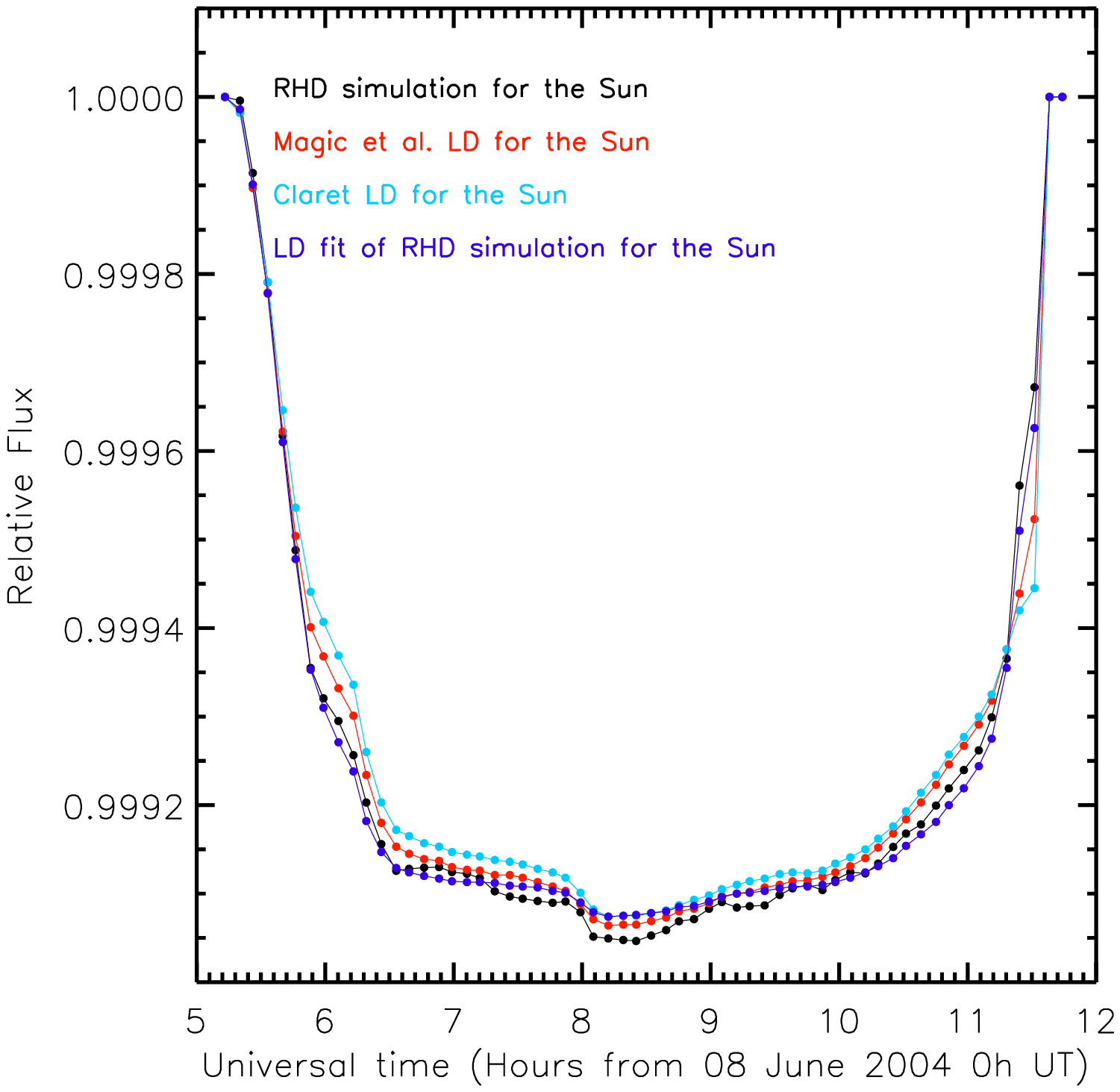}\\
                       \includegraphics[width=0.99\hsize]{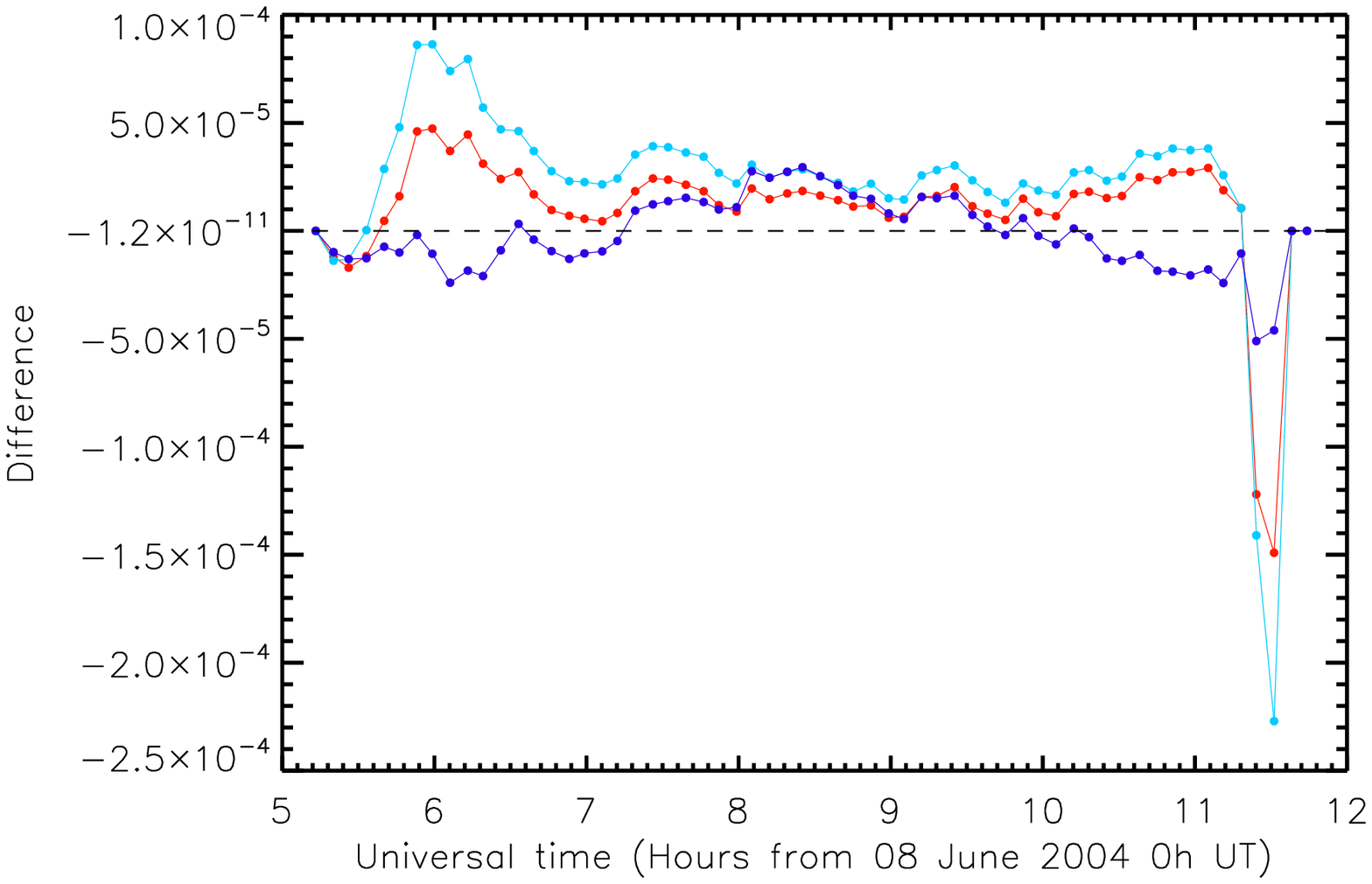}\\
        \end{tabular}
      \caption{\emph{Top panel: }Comparison of different ToV light curves. The black curve is the light curve combining the RHD simulation of the Sun (Fig.~\ref{starplanet} and Fig.~\ref{transit}).  The red and light blue curves represent the limb-darkening \cite{2014arXiv1403.3487M} and \cite{2000A&A...363.1081C}, respectively. In the end, the purple curve represents the limb-darkening carried out using Claret's law for the RHD $<3D>$ profile of Fig.~\ref{multigranulation} (green curve).  In all the cases, a constant value is used for Venus intensity. \emph{Bottom panel: } Difference between the above curves with respect to the transit curve of the RHD simulation of Fig.~\ref{transit}.}
        \label{transitbis} 
   \end{figure}    
   
    \begin{figure}
   \centering
   \begin{tabular}{cc}
                       \includegraphics[width=0.71\hsize]{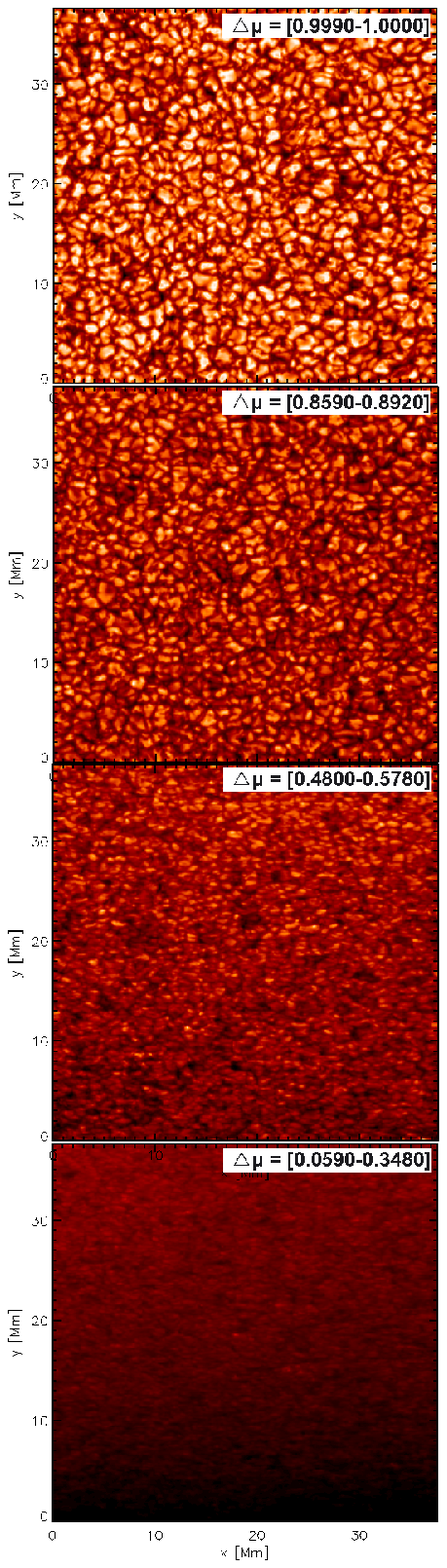}
                               \end{tabular}
      \caption{Sun total intensity image ($370\times370$ pixels) as seen by \textit{HINODE} SOT/SP spectropolarimeter at different $\Delta\mu$ and at continuum wavelengths close to 6300 \AA . Top panel is the center of the Sun and bottom panel is the southern limb. The intensity range is, from the top, $18$--$27\times10^3$ - $17$--$27\times10^3$ - $14$--$21\times10^3$ - $18$--$25\times10^3$\,DN\,pixel$^{-1}$\,s$^{-1}$. 
         }
        \label{hinodeimages1}
   \end{figure}

For the circular subaperture representing the disk of Venus we used a constant flux. Fig.~\ref{transitbis} (top) shows the three different limb-darkening models: \cite{2014arXiv1403.3487M} ($RMSD=9.0\times10^{-5}$, red curve), \cite{2000A&A...363.1081C} ($RMSD=9.3\times10^{-5}$, light blue curve), and the Claret law for the RHD $<3D>$ profile ($RMSD=9.3\times10^{-5}$, purple curve). Fig.~\ref{transitbis} (bottom) displays that the central depression of Venus's transit obtained with the limb-darkened disk of \cite{2000A&A...363.1081C} (light blue curve) is systematically undersized with respect to the transit with the 3D RHD simulation (black curve), while there is a better agreement with \cite{2014arXiv1403.3487M} model (red curve) and with the RHD $<3D>$ limb-darkening. Since the magnitude of the intensity variation in the light curve depends on the brightness contrast between the solar disk and Venus, it is noticeable that RHD synthetic disk image differs from limb-darkening models with particular emphasis at the stellar limb. The contribution of the granulation to the ToV can be retrieved by the absolute difference between the limb-darkening of the RHD $<3D>$ profile and the full RHD ToV (purple curve in bottom panel of Fig.~\ref{transitbis}). We computed the $RMS_{\rm{3D}}=1.7\times10^{-5}$ of this difference and found that it is $\sim$5 smaller than the RMSDs of the different models. Even if the granulation contribution is not large enough to explain the whole observed discrepancies, it is  source of an intrinsic noise due to the stellar variability, that may affect precise measurements of exoplanet transits of, most likely, planets with small diameters.

\subsection{3D RHD granulation versus observed solar granulation}\label{hinodesection}

The aim of this Section is to create a solar disk image using the same procedure of RHD simulation (Section~\ref{tilingsection}) but with, as input, the granulation pattern observed on the Sun. For this purpose, we recovered \textit{HINODE} spectropolarimeter SOT/SP data product collected between December 19 and December 20 2007 from High Altitude Observatory (HAO) website\footnote{http://www.csac.hao.ucar.edu/csac/archive.jsp\#hinode}. The data are reduced and calibrated for all the slit positions (i.e., corresponding to "level1" data). We used SOT/SP Stokes $I$ images (i.e., the total intensity measured) integrated at the continuum wavelengths very close the Fe I line pair at 6300 $\AA$ but avoiding the spectral lines: the emerging intensity is then compatible, in tex of wavelengths probed, to the intensity map of Fig.~\ref{starnoplanet} and the \textit{HINODE} Red filter used in Fig.~\ref{hinode}. The original images of 1024 pixel-spectrograph slit parallel to the North-South polar axis of the Sun was successively located at 20 latitudes and scans of 370 steps of 0.16$\arcsec$ in the direction perpendicular to the slit were performed. At each position in the slit, the images have been integrated for about 5 seconds and for a total time of about 30 minutes (ie., 370 steps). This provides us with 20 (1028$\times$370 pixels) images allowing a complete recording of the center-to-limb variations of the solar disk. However, as the fields of view of the images are partly superimposed we extracted from the \textit{HINODE} images a set of 17 (370$\times$370 pixels) images forming a continuous South-North stripe in the solar southern hemisphere. The solar diameter in that period was 32.5\arcmin ($\sim$975.3\arcsec for the radius), the geometrical distance at the Sun surface, seen under one arcsecond was $R_{\odot \rm{SOHO}}/975.3=713.9$ km.  Each pixel is 0.16$\arcsec$, the size of each image (Fig.~\ref{hinodeimages1}) is then $59.20\times59.20\arcsec$ or $42.18\times42.18$ Mm, with 1 pixel equal to 114 km.

\begin{figure}
   \centering
   \begin{tabular}{c}
                       \includegraphics[width=0.8\hsize]{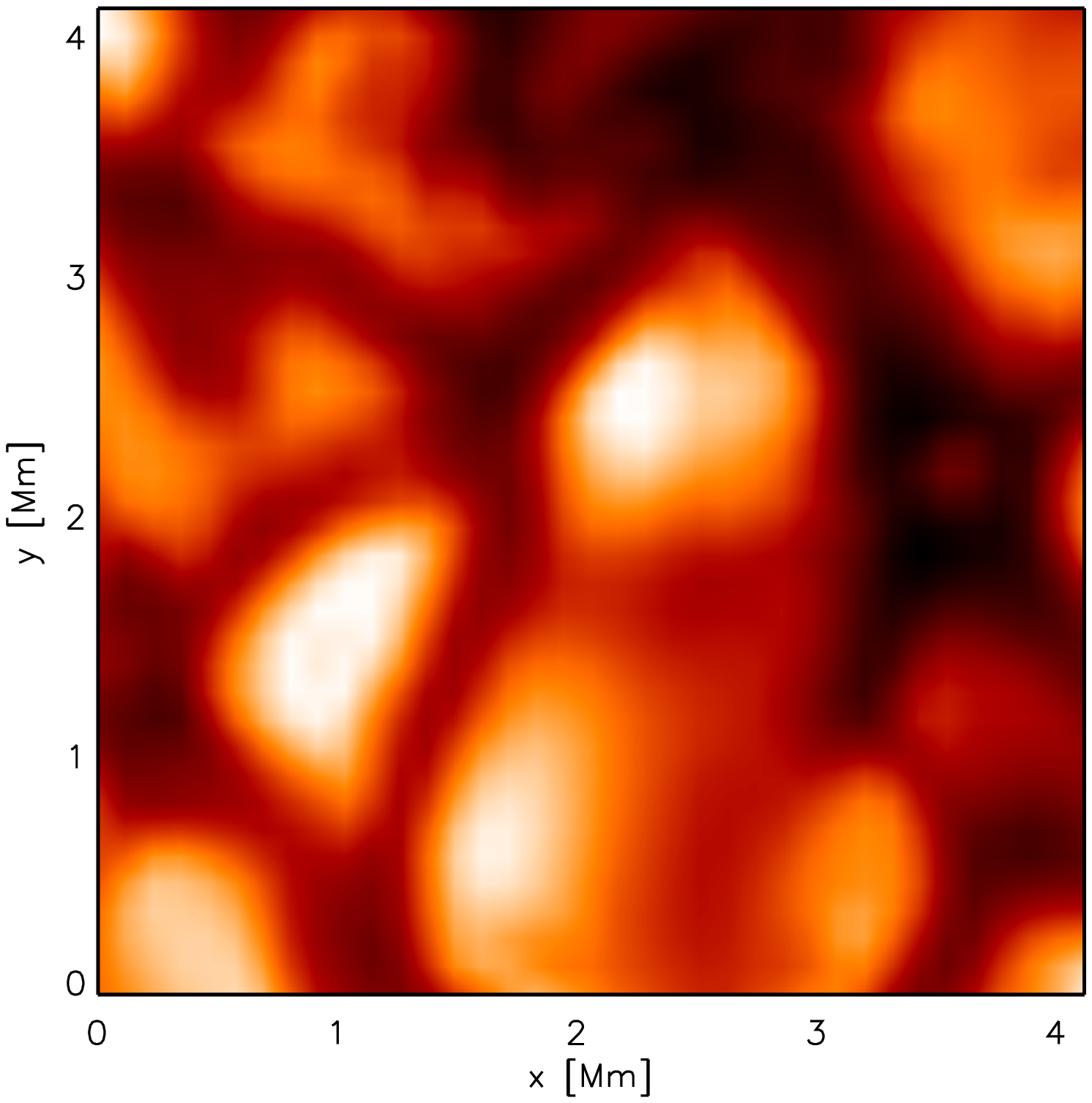}\\
                       \includegraphics[width=0.8\hsize]{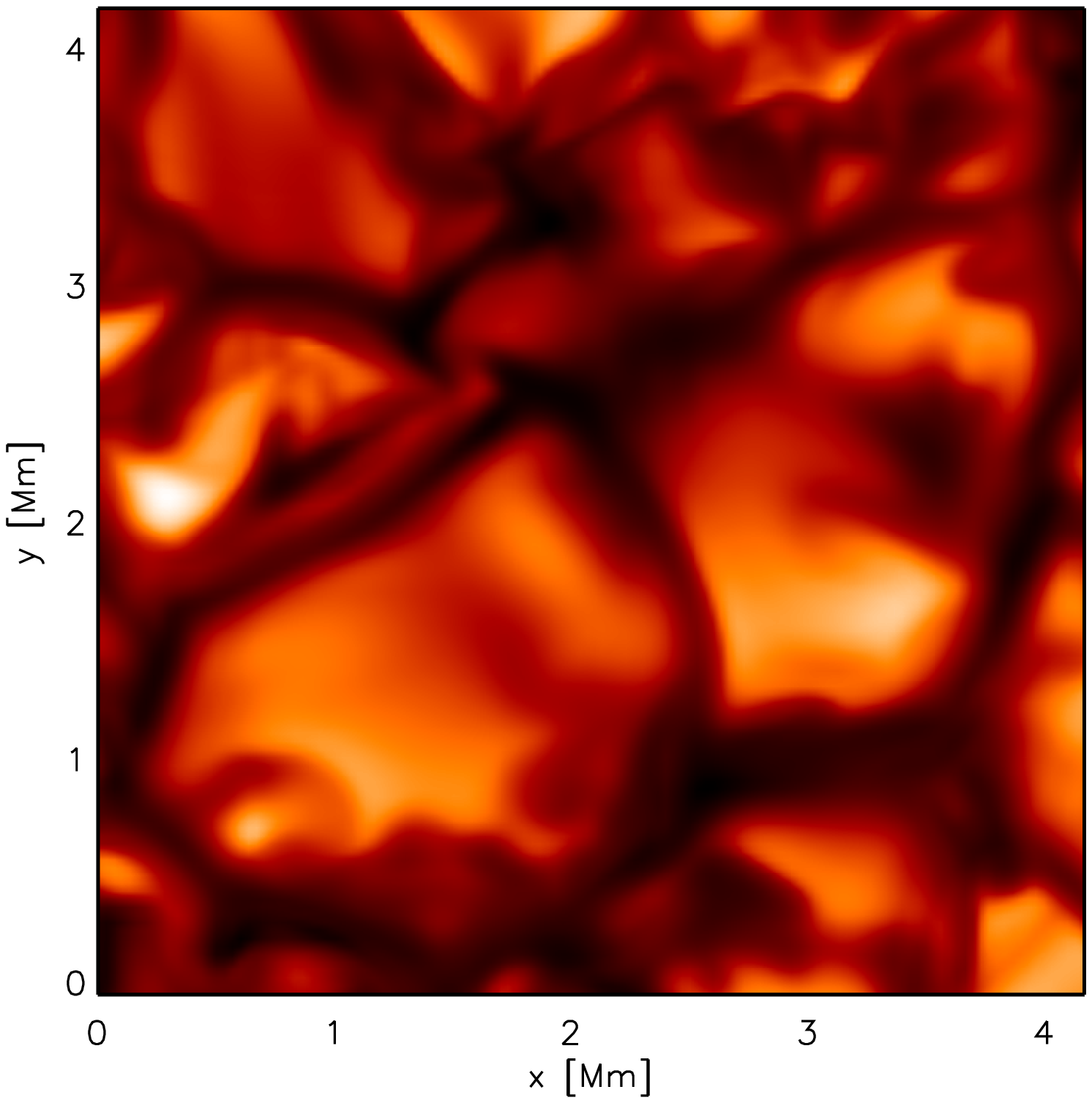}\\
                       \includegraphics[width=0.8\hsize]{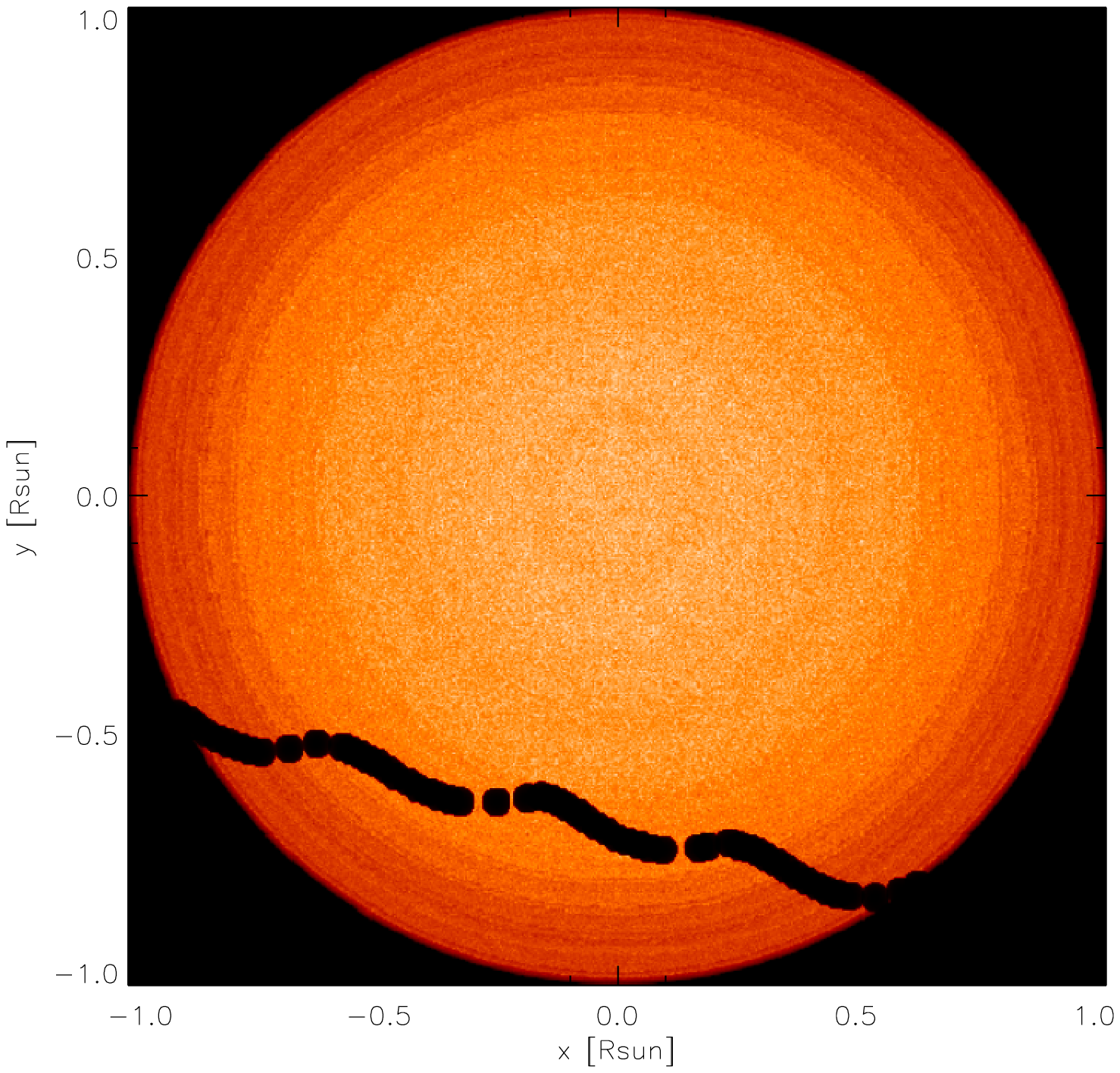}
        \end{tabular}
      \caption{\emph{Top panel:} Cut of $37\times37$ pixels from Fig.~\ref{hinodeimages1} top panel. \emph{Central panel:} intensity map of a small portion of the Sun surface at $\mu$=1 from the RHD of Table~\ref{simus}. The intensity range is [$2.0$--$4.3\times10^6$\,erg\,cm$^{-2}$\,s$^{-1}$\,{\AA}$^{-1}$]. \emph{Bottom panel:} ToV on the solar disk image constructed with 
      the tiling procedure of Section~\ref{tilingsection} but with real granulation observations (top panel and Fig.~\ref{hinodeimages1}) of \textit{HINODE} (see text for details). A constant value is used for Venus intensity.}
        \label{hinodeimages2}
   \end{figure}

\begin{figure}
   \centering
   \begin{tabular}{c}
                       \includegraphics[width=0.99\hsize]{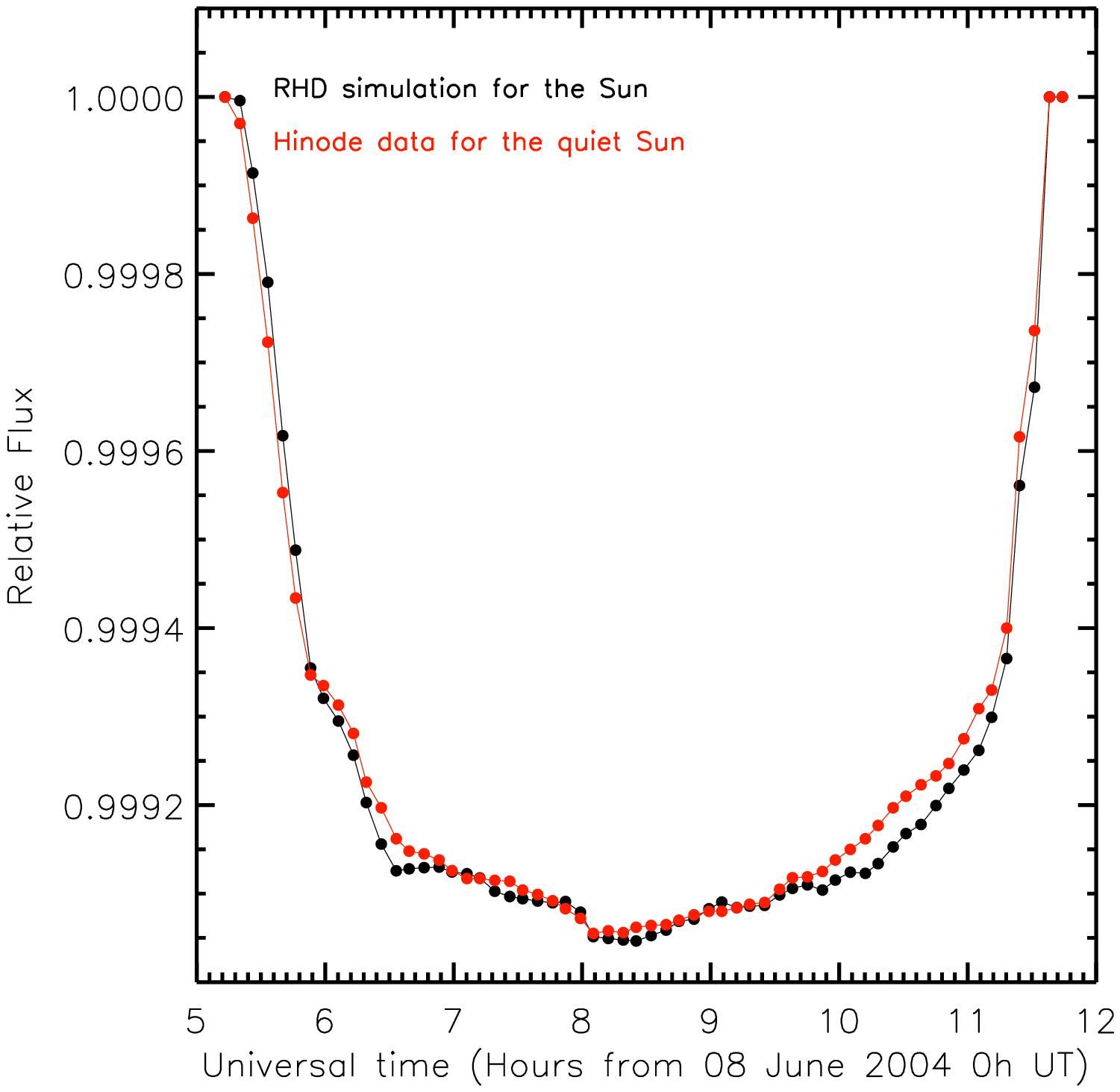}\\
                       \includegraphics[width=0.99\hsize]{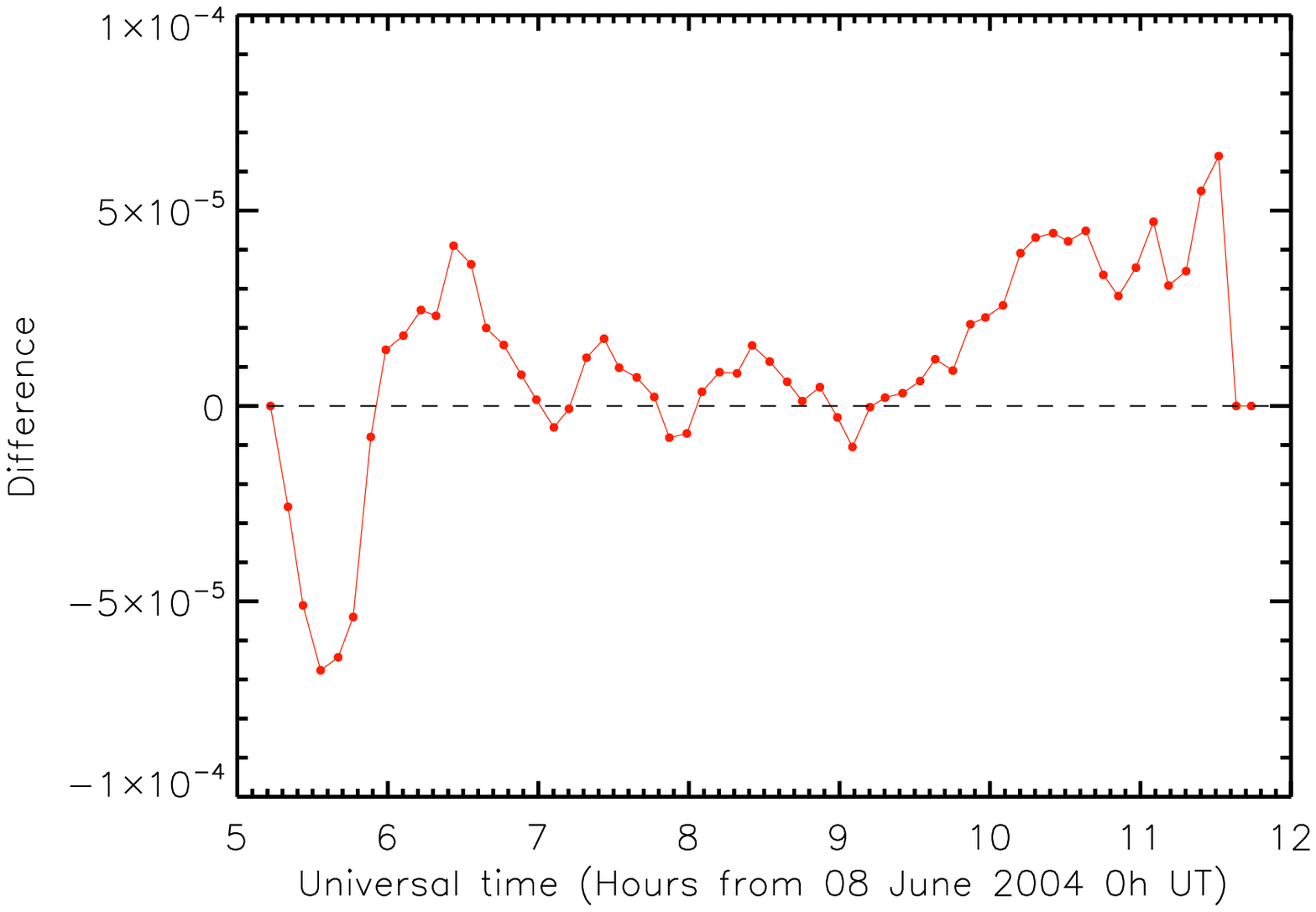}
        \end{tabular}
      \caption{\emph{Top panel: }ToV light curves for RHD simulation (Fig.~\ref{starplanet} and Fig.~\ref{transit}) and \textit{HINODE} SOT/SP spectropolarimeter data (Fig.~\ref{hinodeimages2}, bottom panel). A constant value is used for Venus intensity. \emph{Bottom panel}: Difference between the \textit{HINODE} ToV with respect to the RHD simulation.}
        \label{transitwithhinode}
   \end{figure}

However, the images were too large compared to the synthetic maps (Fig.~\ref{hinodeimages2}, central panel) we used for tiling the spherical surface of the Sun (Fig.~\ref{starnoplanet}) with RHD simulation. We then divided each image by a factor 10 in the East-West axis and by a factor 10 in the North-South polar axis. We ultimately obtained several $37\times37$ images of $\sim$4.2 Mm (Fig.~\ref{hinodeimages2}, top panel) in 83 $\mu$ sub-steps.\\
In the end, we applied the tiling procedure of Section~\ref{tilingsection} and obtained the ToV for a granulation pattern based on \textit{HINODE} observations (Fig.~\ref{hinodeimages2}, bottom panel). The deviation from \textit{ACRIMSAT} observations is $RMSD=8.7\times10^{-5}$. Fig.~\ref{transitwithhinode} displays the comparison between the light curve obtained from the RHD simulation (Fig.~\ref{transit}, black dashed line) and the one calculated from the ToV of Fig.~\ref{hinodeimages2} (bottom panel). The absolute difference between the ToV curves (Fig.~ \ref{transitwithhinode}, bottom panel) shows a typical value of $\sim5.0\times10^{-5}$, which is close to what is reported in Fig.~\ref{transitbis} (bottom). However, the agreement is better, in particular for the Ingress/Egress slopes of the transit. This confirms that the limb-darkening and the granulation pattern simulated in the 3D RHD solar simulation represent well what is imaged by \textit{HINODE}. This comparison with \textit{HINODE} data is important because allows another inspection of RHD simulation using data with smaller noise with respect to ToV data.\\
Following what we did in Section~\ref{limbdarkeningsection} for the RHD <3D> profile, we used the limb-darkening law of Claret to fit the intensity profile obtained from \textit{HINODE} data (Fig.~\ref{profilecomparison}, green curve). The coefficients are reported in Table~\ref{coeff}. It is also noticeable in Fig.~\ref{profilecomparison} the good agreement between the Hinode and the RHD profiles. To obtain the contribution from the granulation, we computed the limb-darkening ToV and subtracted it to the ToV of the \textit{HINODE} image. The resulting signature of the granulation has an $RMS_{\rm{Hinode}}=3.0\times10^{-5}$, which is comparable to the $RMS_{\rm{3D}}=1.7\times10^{-5}$ due to the granulation signal in RHD ToV. Both values are however lower than or of about the same order of the smallest error bar of \textit{ACRIMSAT} ($2.7\times10^{-5}$), and thus the granulation signal cannot be unveiled with these data. Nevertheless, It should be note that there is a possibility that the small scale variability produced by the RHD simulation may be slightly underestimated due to the snapshot chosen.

\section{Conclusions and implications for transit studies of extrasolar planets}

Modeling the transit light curve of Venus as seen from \textit{ACRIMSAT} implies the importance to have a good representation of the background solar disk and of the planet itself. For the Sun, we used the realistic, state-of-the-art, time-dependent, radiative-hydrodynamic stellar atmosphere of the Sun from the \textsc{Stagger}-grid. For Venus, we extracted and fitted radially averaged profiles from high spatial resolution images of \textit{HINODE}.

We provided synthetic solar disk image using the spherical tile imaging method already applied in \cite{2014A&A...567A.115C,2012A&A...540A...5C,2010A&A...524A..93C}. We applied a statistical approach to show that our RHD simulation of the Sun is adapted (in terms of limb-darkening and emerging flux) to interpret such a data and that the granulation fluctuations have an important effect of the light curves during the transit. Our procedure successfully explain \textit{ACRIMSAT} observations of 2004 ToV and showed that the granulation pattern causes fluctuations in the transit light curve. We compared different limb-darkening models to our RHD simulation and evaluated the contribution of the granulation to the ToV. We showed that the granulation pattern can partially explain the observed discrepancies between models and data. However, the amplitude of the signal is close to the typical differences between the different models used and it is difficult to disentangle the signal of the granulation with the data used in this work. Moreover, we found that the Venus's aureole contribution during ToV is $\sim10^{-6}$ times less intense than the solar photosphere, and thus, accurate measurements of this phenomena are extremely challenging.

In the end, we applied the same spherical tile imaging procedure to the observations of center-to-limb Sun granulation with \textit{HINODE} and built a solar disk image. We then compared the resulting transit curves and found that the overall agreement between real and RHD solar granulation is very good, either in term of depth or Ingress/Egress slopes of the transit. This confirms that the limb-darkening and the granulation pattern simulated in 3D RHD Sun represent well what is imaged by \textit{HINODE}. 
 
3D RHD simulations are now established as realistic descriptions for the convective photospheres of various classes of stars. They have been recently employed: (i) in the prediction of differential spectroscopy during exoplanet transits to reconstruct spectra of small stellar surface portions that successively become hidden behind the planet \citep{2014arXiv1408.1402D}; (ii) or to assess the 
transit light curves effect of the limb-darkening issued from 3D RHD simulations with respect to 1D models. \\
The ToV is an important benchmark for current theoretical modeling of exoplanet transits. Being able to explain consistently the data of 2004 transit is then a new step forward for 3D RHD simulations. This implies that our tiling procedure can be adapted to exoplanet transits. Moreover, our statistical approach in this work allowed to show that the granulation have to be considered as an intrinsic incertitude, due to the stellar variability, on precise measurements of exoplanet transits of, most likely, planets with small diameters. A possible solution to reduce this incertitude is the repetition of the transit measurements several times rather than take a single snapshot.

The prospects for planet detection and characterization with transiting methods are excellent with access to a large amount of data for different kind of stars either with ground-based telescopes such as HATNet\citep[Hungarian Automated Telescope Network,][]{2004PASP..116..266B}, NGTS \citep[Next Generation Transit Survey,][]{2013EPJWC..4713002W}, TRAPPIST \citep[TRAnsiting Planets and PlanetesImals Small Telescope,][]{2011Msngr.145....2J}, WASP \citep[Wide Angle Search for Planets, ][]{2006PASP..118.1407P} or space-based missions like CHEOPS \citep[CHaracterizing ExOPlanet Satellite, ][]{2013EPJWC..4703005B,2010ApJ...713L..79K}, COROT \citep[Convection, Rotation and planetary Transits,][]{2006cosp...36.3749B,2006ESASP1306...33B}, Kepler \citep{2010Sci...327..977B}, TESS \citep[Transit Exoplanet Survey Satellite,][]{2010AAS...21545006R}, PLATO \citep[PLAnetary Transits and Oscillation of stars,][]{2014ExA...tmp...41R}. \\
In this context, 3D RHD simulations are essential for a detailed quantitative analysis of the transits. Indeed, the interpretation and the study of the impact of stellar granulation is not limited to the Sun \citep{2009LRSP....6....2N} but RHD simulations cover a substantial portion of the Hertzsprung-Russell diagram \citep{2013A&A...557A..26M,2013ApJ...769...18T,2009MmSAI..80..711L}, including the evolutionary phases from the main sequence over the turnoff up to the red-giant branch for low-mass stars.

\begin{acknowledgements}   
The authors thank G. Schneider, J. M. Pasachoff, R. C. Willson for providing 2004 ToV data and for the enlightening discussions. The authors acknowledge Alphonse C. Sterling (NASA/Marshall Space Flight Center) for the preparation and diffusion of the transit images. CP acknowledges funding from the European Union Seventh Framework Program (FP7) under grant agreement number 606798 (EuroVenus). RC is the recipient of an Australian Research Council Discovery Early Career Researcher Award (project number DE120102940). 
\end{acknowledgements}


   \bibliographystyle{aa}
\bibliography{biblio.bib}

\end{document}